**Laboratory Studies of Methane and its Relationship to Prebiotic Chemistry**

Kensei Kobayashi[1], Wolf D. Geppert[2], Nathalie Carrasco[3], Nils G. Holm[2], Olivier Mousis[4],

Maria Elisabetta Palumbo[5], J. Hunter Waite[6], Naoki Watanabe[7], Lucy M. Ziurys[8]

[1]Department of Chemistry, Yokohama National University, Hodogaya-ku, Yokohama 240-8501, Japan

[2]Department of Geological Sciences, Stockholm University, Svante Arrhenius väg 8, SE-106 91 Sweden

[3]LATMOS, Université Versailles St-Quentin, UPMC, CNRS, 11 bvd d'Alembert,78280 Guyancourt, France

[4]Aix Marseille Université, CNRS, LAM (Laboratoire d'Astrophysique de Marseille) UMR 7326, 13388, Marseille, France

[5]INAF - Osservatorio Astrofisico di Catania, via Santa Sofia 78, 95123, Catania, Italy

[6]Southwest Research Institute, San Antonio, Texas 78228-0510, USA

[7]Institute of Low Temperature Science, Hokkaido University, Kita-19, Nishi-8, Kita-ku, Sapporo 060-0819, Japan

[8]Department of Astronomy, Department of Chemistry and Biochemistry, and Steward Observatory, University of Arizona, 933 North Cherry Avenue, Rm. N204  Tucson, AZ 85721-0065, USA

**Abstract**

To examine how prebiotic chemical evolution took place on Earth prior to the emergence of life, laboratory experiments have been conducted since the 1950s.   Methane has been one of the key molecules in these investigations.   In earlier studies, strongly reducing gas mixtures containing methane and ammonia were used to simulate possible reactions in the primitive atmosphere of Earth, producing amino acids and other organic compounds. Since Earth's early atmosphere is now considered to be less reducing, the contribution of extraterrestrial organics to chemical evolution has taken on an important role   Such organic molecules may have come from molecular clouds and regions of star formation that created protoplanetary disks, planets, asteroids, and comets. The interstellar origin of organics has been examined both experimentally and theoretically, including laboratory investigations that simulate interstellar molecular reactions. Endogenous and exogenous organics could also have been supplied to the primitive ocean, making submarine hydrothermal systems plausible sites of the generation of life. Experiments that simulate such hydrothermal systems where methane played an important role have consequently been conducted. Processes that occur in other solar system bodies offer clues to the prebiotic chemistry of Earth. Titan and other icy bodies, where methane plays significant roles, are especially good targets. In the case of Titan, methane is both in the atmosphere and in liquidospheres that are composed of methane and other hydrocarbons, and these have been studied in simulation experiments. Here, we review the wide range of experimental work in which these various terrestrial and extraterrestrial environments have been modeled, and we examine the possible role of methane in chemical evolution.

## 1. Introduction

How life on Earth emerged is one of the most important questions that have been left unanswered. In the 1920s, Oparin and Haldane independently published their theories on the generation of life (Deamer and Fleischaker, 1994). The Oparin-Haldane Theory has been referred to as the chemical evolution hypothesis, where life was born after the evolution of organic materials that were dissolved in the primitive ocean. The processes of chemical evolution were, however, considered too difficult to be examined by experiments in the early 20th century.

Miller (1953) reported the first successful abiotic synthesis of amino acids by spark discharges in a gas mixture of methane, ammonia, hydrogen, and water, which showed that the chemical evolution hypotheses could be tested. From the 1950s to 1970s, a large number of experiments that simulated reactions in the primitive Earth atmosphere were conducted (Miller and Orgel, 1974). In most of these studies, strongly reducing gas mixtures containing methane and ammonia as major constituents were used as starting materials. In addition to spark discharges, energy sources such as ultraviolet light (Sagan and Khare, 1971), thermal energy (Harada and Fox, 1964), ionizing radiation (Ponnamperuma et al., 1963), and shock waves (Bar-Nun et al., 1970) were used, and successful syntheses of amino acids and/or nucleic acid bases were reported.

Planetary explorations in the late 20[th] century offered novel information on the formation of the solar system, including Earth, which suggested that the primitive Earth atmosphere was never strongly, but only slightly, reducing (Matsui and Abe, 1986a, b; Kasting, 1990). Major gases in this case would have been carbon dioxide and nitrogen, together with some reducing gases such as carbon monoxide and hydrogen as minor components (Kasting, 1993). If this were the case, prebiotic synthesis of amino acids and other bioorganic compounds would have been quite difficult. For example, only small amounts of amino acids were formed when a gas mixture of carbon dioxide (or carbon monoxide), nitrogen, and water was subjected to spark discharges unless high a concentration of hydrogen was added to the mixture (Schlesinger and Miller, 1983). However, amino acid formation from atmospheres without methane and ammonia could still have been possible if high-energy particle irradiation from cosmic rays and high temperature plasmas, which simulated bolide impact of meteorites, were used. Production of amino acid precursors and nucleic acid bases was reported from a mixture of carbon dioxide, carbon monoxide, nitrogen, and water with such particle bombardment (Kobayashi et al., 1990, 1998; Miyakawa et al., 2002).

On the other hand, a wide variety of organic compounds have been detected in extraterrestrial environments. Interstellar organic molecules have been identified in dense clouds by infrared and/or microwave telescopes for many decades (Herbst and van Dishoeck, 2009), and organic molecules were directly detected with mass spectrometers onboard spacecraft in bodies such as comets (Goesmann et al. 2015), Jupiter (Nixon et al., 2007; Sada et al., 2007), and Titan (Vuitton et al., 2007; Cravens et al., 2006). Extracts from carbonaceous chondrites have been comprehensively analyzed, and a great number of amino acids and some nucleobases were identified together with hydrocarbons, carbonic acids, amines, and other organic compounds (Martins et al. 2008; Botta et al. 2008). These extraterrestrial organic compounds have been considered to be important building materials for the first life on Earth. How were organic compounds found in comets and meteorites formed? One of the plausible sites is the interstellar medium. As will be discussed in Section 2, a variety of experiments simulating reactions in interstellar environments have been conducted in which organic compounds were formed.

After delivery of extraterrestrial organic compounds by asteroids, meteorites, comets, and/or interplanetary dust particles (IDPs) to primitive Earth, further chemical evolution towards the generation of early life probably took place in the primeval ocean. The perception of the primitive ocean has drastically changed after the discovery of submarine hydrothermal vents in the late 1970s. Submarine hydrothermal systems (SHTSs) have been considered sites for generation of early life on earth for several reasons (Hennet *et al.*, 1992). One of them is that the SHTSs could maintain reducing environments even on present Earth, since reducing gases including methane, ammonia, hydrogen, and hydrogen sulfide are quite rich in the hydrothermal fluids. Another reason is that the lineages of the hyperthermophilic chemoautotrophs that thrive in such vents are relatively close to the hypothetical position of the last common universal ancestor (LUCA) on the 16S rRNA tree of life (Pace, 1991).

In Section 3, laboratory experiments simulating SHTS environments are discussed. Two types of experiments have been conducted: those with closed systems by the use of autoclaves, and those with flow systems. A wide variety of organic compounds have been abiotically synthesized from molecules found in SHTSs such as methane. Possible chemical evolution from amino acids or their precursors to peptide and organic aggregates/globules in SHTSs has been suggested.

Though methane is no longer regarded as a plausible major constituent in a primitive Earth atmosphere, it could have been a minor component in it. Titan, the largest Kronian satellite, has a dense atmosphere, where nitrogen is the dominating species (more than 90 %) followed by methane (a few %). Titan is called a natural laboratory of chemical evolution, where various reactions between nitrogen and methane are taking place triggered by solar ultraviolet light and the impact of magnetospheric electrons, meteor impacts, and cosmic radiation. In addition, lakes of liquid hydrocarbons have been found on the surface of Titan. Chemical evolution in hydrocarbon solvents, not in water, has been also examined by simulation experiments. In Section 4, we will review such simulation experiments and discuss the possible generation of life on Titan and its relevance to that on primitive Earth.

In the final Section, we discuss the significance of simulation experiments as a guide for possible reaction pathways toward the generation of life on Earth, and map out future prospects.

## 2. Experiments on reactions involving methane in the interstellar medium

Methane has no permanent electric dipole moment because of its tetrahedral symmetry. Therefore, it cannot be studied via radiofrequency observations of its pure rotational transitions, the most general method to study interstellar molecules. Instead, it can be traced by observing its vibrational transitions in the infrared region of the electromagnetic spectrum (e.g., Knez et al. 2009). Interstellar excitation of vibrational levels of molecules like methane usually requires a strong background source that emits in the infrared, such as a very young or very old star. Therefore, observational studies of methane have been principally limited to the envelopes of dying stars such as IRC+10216 or CRL 618 (e.g. Cernicharo et al. 2001) and in dense molecular clouds with young, imbedded protostars (e.g. Boogert et al. 2004), including a few with protoplanetary disks (e.g. Gibb & Horne 2013). Detections of interstellar methane in fact are rather limited. Methane is clearly created from gas-phase, equilibrium chemistry in old stars, where the high densities and temperatures near the stellar photosphere promotes its formation (e.g., Charnley et al. 1995). Its origin in molecular clouds, on the other hand, may be through gas-phase reactions or on surface processes on dust grains, or both. Because the abundance and distribution of methane is not well documented due to observational limitations, it is very important to understand its role and connection to other hydrocarbons that are more readily studied in the interstellar medium, such as $C_2H$ or CCH, which have permanent electric dipole moments. Molecules such as these species are fairly universally found in interstellar sources, including planetary nebulae, diffuse clouds, and cold, dark clouds (e.g., Ziurys et al. 2015). They therefore serve as "proxy" molecules for methane, and relating their reaction pathways to those of methane, including photo-detachment of negative ions, is critical to obtain a complete picture of where $CH_4$ exits and its abundance.

## 2.1. Experimental studies into gas phase methane and hydrocarbon chemistry

Carbon-containing molecules, including methane and other hydrocarbons, play a decisive role in the chemistry of the interstellar medium (ISM) and the atmospheres of planets and their satellites. Also, hydrocarbons, including methane, are key components of the atmospheres of gaseous planets and their satellites (Sakai *et al.*, 2009), especially Titan, which has a methane cycle similar to the terrestrial water cycle and contains hydrocarbon lakes on its surface. Many of the interstellar carbon-bearing molecules are thought to be formed by gas-phase processes. Such processes are important in prebiotic syntheses, as evidenced by the high degree of deuterium enrichment found in organic molecules extracted from meteorites, and in the insoluble organic material (IOM) (e.g., Pizzarello et al. (2006)). High D-enhancements cannot be accounted for easily by anything other than low temperature, gas-phase reactions (Millar 2005). Therefore, it is important to understand gas-phase reactions in the ISM.

Due to the low temperatures prevalent in the interstellar medium, neutral-neutral reactions between closed shell molecules are highly unlikely, since these processes usually have an activation energy barrier that usually cannot be surmounted under interstellar conditions. Radical- and ion-induced reactions and processes on grain surfaces be the mechanisms for molecule production. Furthermore, the low particle density in interstellar objects prevents three-body reactions. Nevertheless, a rich carbon chemistry has been detected in a multitude of interstellar objects like dark clouds and star-forming regions. The main pathways in this chemistry can be grouped under the following categories: (1) ion induced reactions, (2) radical- and atom-induced reactions, (3) photoprocesses, (4) ion-electron, and (5) ion-ion reactions. Examples are as follows:

*Ion-neutral reactions*

Radiative association:    $CH_3^+ + H_2 \rightarrow CH_5^+ + h\nu$ (i)

Proton transfer:    $CH_5^+ + CO \rightarrow CH_4 + HCO^+$ (ii)

Neutral transfer:    $H_2 + CH^+ \rightarrow CH_2^+ + H$ (iii)

*Radical- and atom-induced reactions*

Neutral-neutral processes:    $C + C_2H_2 \rightarrow C_3H + H$ (iv)

*Photon-induced processes*

Photoionisation:    $CH_4 + h\nu \rightarrow CH_4^+ + e^-$ (v)

Photodetachment:    $C_4H^- + h\nu \rightarrow C_4H + e^-$ (vi)

*Ion-electron reactions:*

Dissociative recombination:    $CH_5^+ + e^- \rightarrow CH_4 + H$ (vii)

*Ion-electron reactions:*

Mutual neutralisation    $H^+ + H^- \rightarrow H + H$ (viii)

Different reaction pathways, including some of the above-mentioned processes, have been postulated to form methane, hydrocarbons, and other C-bearing molecules. In the case of the methane, a generally accepted and unambiguous formation pathway is still missing. To identify feasible gas-phase pathways to form molecules in the ISM, knowledge of the rate constants and product branching ratios leading to these species is vital. The prediction of these parameters by theoretical calculations has been proven difficult for many reactions, especially dissociative recombination (Kokoouline *et al.*, 2011), and experimental investigation of interstellar reactions faces severe challenges. Ideally, experiments should be carried out under conditions (pressure, temperature, excitation state of the reactants) similar to those encountered in the ISM. In many experimental devices, low pressures in the range of $10^{-11}$ mbar can be achieved, which are, however, still several orders of magnitude higher than those encountered in interstellar environments. Nevertheless, for most gas-phase studies it is crucial to exclude three-body processes, which is accomplished under these pressures (Geppert and Larsson, 2013). Matching the low temperatures in the ISM can be a bit more of an issue. Many devices capable of investigating interstellar hydrocarbon reactions are fairly large and thus difficult to cool efficiently. Nevertheless, cooled storage rings have recently been constructed in Heidelberg (von Hahn *et al.*, 2016; Zaifman *et al.*, 2005) and Stockholm (Schmidt *et al.*, 2013). Another issue is the "freeze-out" of reactants on the inner surfaces of the experimental apparatus at very low temperatures. Also, radical and ionic reactants are often produced in discharge devices. This creates species often in highly excited states. In a high vacuum, deactivation of these states can only happen by photon emission, and lifetimes for these decay processes might be high. In some set-ups like ion traps cooling by cold buffer gases before the experiment has been employed successfully (Gerlich and Borodi, 2009). Also supersonic expansions have been used to cool reactants (McCall *et al.*, 2004).

Also, many of the reactants in interstellar chemistry are difficult to produce without by-products. For ions, mass spectrometers, and deflection magnetic fields can be used to separate the desired reactant from contaminants (which only works if they have different masses, so different isomers and isobars cannot be separated). In case of radicals, production of pure, cool reactants can be difficult. However, hydrogen atoms generated by discharges can be equilibrated to cool temperatures by flowing through cold accommodator nozzles (Gerlich *et al.*, 2012). We will now briefly discuss the different experimental methods to investigate the above-mentioned processes.

For investigations of ion-neutral reactions, selected ion flow tubes (SIFT) and ion traps have been proven successful. A scheme of the selected ion flow tube is depicted in Fig. 1.

FIG. 1. Scheme of a selected ion flow tube apparatus.

In this technique, the ions produced in the source are mass-selected by a quadrupole mass filter. The carrier gas entered into the tube moves the ions to the reaction zone, in which the reaction gases are added. Product ions formed by the ion−neutral reactions are mass selected and detected by a channeltron. A multitude of ion-neutral reactions involving hydrocarbons has been studied with SIFT (Anicich *et al.*, 2004).

Also radiofrequency ion traps like the 22-pole ion trap developed by Gerlich and co-workers (Gerlich, 2003) are used to investigate ion-neutral reactions. In such devices, the ions, which have previously been mass selected, can be cylindrically trapped by an electrical field changing on a time scale comparable to radiofrequencies. The axial trapping is performed by charged end caps. During the trapping time, ions can be cooled by cold buffer gases, whereupon the reactant gas is added to the ion trap. After a certain reaction time the ions are axially ejected from the trap by an electrical field. The decay of the reactant ion and the build-up of product ions can be followed through detection of these ions after different reaction times. For example, the rate constant of the reaction

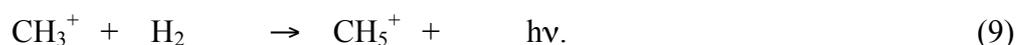

$$CH_3^+ \; + \; H_2 \qquad \rightarrow \; CH_5^+ \; + \qquad h\nu. \qquad\qquad (9)$$

was measured in an ion trap as $6 \times 10^{-15}$ cm$^3$ s$^{-1}$ at 80 K by Gerlich and Kaefer (1989). The obtained value is almost 50 times the value predicted by previous theoretical calculations (Bates, 1987). Nevertheless, due to the high amount of $H_2$ present in dark clouds, the process is still feasible.

Ion traps can also be employed to investigate photodetachment of anions. The absolute cross sections of the anions of the photodetachment of the interstellar hydrocarbon anions $C_4H^-$ and $C_6H^-$ as well as $C_3N^-$ and $C_5N^-$ could recently be determined in a study using ion traps (Best et al., 2011, Kumar *et al.* 2013). Such photodetachment reactions, which, in the case of $C_4H^-$ and $C_6H^-$, lead to neutral hydrocarbons that can further be involved in the hydrocarbon chemistry of the interstellar medium and planetary atmospheres. These processes can play a pivotal role in photon-dominated regions of dark clouds, from which the abundances of molecular anions can be inferred. Even in the inner parts of dark clouds in which UV photons cannot penetrate, UV photodetachment can still constitute a route of destruction of anions because secondary electrons produced by cosmic rays can excite molecules to high Rydberg states, which emit UV radiation upon decay. Furthermore, photodetachment of negative ions is one of the major destruction mechanisms in circumstellar envelopes. Model calculations of the circumstellar envelope IRC+10216 predicted that, in regions where anions display their peak abundance, photodetachment constitutes as the most efficient destruction mechanism of these species (Kumar *et al.* 2013). Furthermore, photodetachment is an important destruction process of anions in the upper layers of Titan's ionosphere (Vuitton et al. 2009).

FIG. 2. Schematics of a flowing afterglow apparatus

Ion–electron reactions can be studied with flowing afterglow devices, merged beams, and storage rings. In the ISM, ion–electron reactions of molecules are almost restricted to exclusively dissociative recombination, since possibly competing processes are either too slow (radiative recombination) or highly endoergic (e.g., dissociative ionization). Many dissociative recombination reactions have been studied with the flowing afterglow technique (Schmeltekopf and Broida, 1963; Ferguson et al., 1964) , which is a further development of the earlier stationary afterglow method and has the advantage that the environments of ion production and reaction are spatially separated. In a flowing afterglow apparatus (see Fig. 2) usually $He^+$ ions are created by a microwave discharge. Downstream in the gas flow the reactant ions are produced through reactions of appropriate precursor molecules with $He^+$, sometimes via a sequence of several reactions. Further on the reactive gas is added, and the decay of the probe ions can be followed by a mass spectrometer, since the distance between the reagent inlet and the mass spectrometer can be varied. A movable Langmuir probe can serve to follow the decay of the electrons. This Flowing Afterglow Langmuir probe (FALP) technique (Smith *et al.*, 1984) is used to determine the rate and, in some cases, the products of dissociative recombination.

Furthermore, magnetic storage rings have been used to determine the rates and branching ratios of dissociative recombination reactions. As an example the CRYRING storage ring which was located at Stockholm University is depicted in Fig. 3. In a storage ring, ions are usually produced in a hollow cathode discharge or a tandem accelerator. After being mass selected by a bending magnet, they are injected into the ring. Thereafter, they are stored for several seconds to allow for radiative cooling. In a segment of the ring called the electron cooler, they are then merged with a cold electron beam at low relative kinetic energies down to several meV or even lower. The neutral fragments leave the ring tangentially and can then be detected. The employment of the grid method (Larsson and Thomas, 2001) or energy-sensitive multi-strip surface-barrier (EMU) detectors (Buhr *et al*., 2010) allows determination of the branching fractions between different reaction pathways. As mentioned, dissociative recombination reactions are highly exoergic and can usually show 2 or more different energetically feasible reaction channels. Also, break-up of the ion into 3 fragments upon dissociative recombination is much more common than is predicted by previous theoretical studies (Bates, 1986). This is, for instance, the case in the dissociative recombination of protonated methane:

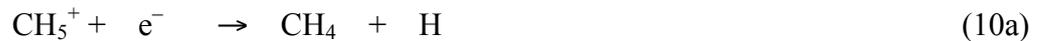
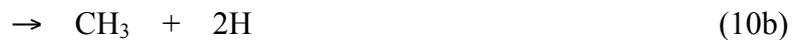
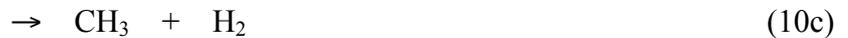
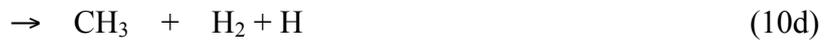

$$CH_5^+ + e^- \rightarrow CH_4 + H \tag{10a}$$
$$\rightarrow CH_3 + 2H \tag{10b}$$
$$\rightarrow CH_3 + H_2 \tag{10c}$$
$$\rightarrow CH_3 + H_2 + H \tag{10d}$$

in which Reaction (10a) represents only 5 % of all recombination processes and reaction 10b is the major pathway (Semaniak *et al*., 1998). This fact has raised questions about the feasibility of the former process as the final step in the formation of methane in the ISM (Viti *et al*., 2000). Hydrogenation of carbon by H atoms on interstellar grain surfaces has been discussed as an alternative mechanism (Bar-Nun *et al*., 1980).

In some cases, there have been differences between rate constants measured by flowing afterglows and in storage rings. For the dissociative recombination, the storage ring measurements (Smith *et al.*, 1984) yielded a rate constant of 2.8 $(T/300)^{-0.52} \times 10^{-7}$ cm$^3$ s$^{-1}$, whereas afterglow experiments reported considerably higher values amounting up to $1.1 \times 10^{-6}$ cm$^3$ s$^{-1}$ at 300 K (Adams *et al.*, 1984). This discrepancy can be caused by different reasons. In afterglow studies, where usually higher pressures pertain, three-body reactions involving the buffer gas or another reactant molecule or atom must be taken into account. Also, highly excited intermediate neutral molecules formed by the attachment of the electron or highly excited neutral fragments can be cooled by collisions with buffer gases and consequently follow a different reaction pathway. On the other hand, ions in excited states that show different behaviour in dissociative recombination from those in the ground states could be present in the storage ring experiments.

FIG. 3. The CRYRING storage ring

Crossed-beam experiments have proven to be powerful tools for investigating neutral-neutral reactions. In these devices, two supersonic beams (one containing the atom or radical and one the other reactant) are crossed. Often the intersection angle can be scanned in these devices enabling continuous variation of the relative kinetic energy of the reactants. Detection of the products can be accomplished with laser-induced fluorescence and mass spectrometry. The reaction of carbon at very low translational energy of the reaction of carbon atoms with unsaturated hydrocarbons, for example,

$$C_2H_4 + \quad C \quad \rightarrow \quad C_3H_3 \quad + \quad H \qquad (11)$$

which could be an important step in the build-up of larger hydrocarbons and could be measured at relative translational energies down to 0.5 kJ/mol (Geppert *et al.*, 2000). With so-called "soft ionization" of product species by electrons with low energies and subsequent detection of the ions by mass spectrometry, even different products of the atom-neutral reaction could be identified (Balucani *et al.*, 2006).

For determining the rate constant of atom-neutral reactions, the CRESU (Cinetique des Reactions en Ecoulement Supersonique Uniforme) apparatus has been employed with great success for studying hydrocarbon reactions at ISM temperatures (Canosa *et al.*, 1997).  In the CRESU, these cold temperatures are achieved by expanding the reactants and the radical precursors in buffer gas through a supersonic flow through a convergent-divergent Laval nozzle, whereby the buffer gas makes up the vast majority (~99 %) of the gas mixture.  Reactant atoms and radicals can then be produced from the precursor through irradiation of a coaxial laser.

FIG. 4. Schematic of the CRESU apparatus.

The decay of the reactant atom (e.g., C($^3$P)) can then be followed by laser-induced fluorescence since the distance between the movable reservoir and the detection zone can be continuously altered and the velocity of the gas flow is well defined. Pseudo-first order conditions (excess of the neutral reactant gases, e.g., acetylene relative to the reacting atoms or radicals) are usually employed, which enables measurement of the rate constants. Very low temperatures (down to 15 K) can be routinely achieved in the CRESU device. Also the supersonic flow acts as a wall-free reaction vessel excluding problems arising from reaction on the chamber walls, which can be a problem in flow tubes. As can be seen in Fig. 4., the CRESU device can also be used for ion-neutral reactions (Rowe *et al.*, 1984).

Unfortunately, there is not very much data available about cation-anion reactions (e.g., mutual neutralization processes), since these are very demanding to study experimentally. Investigations of ion-ion reactions require production and mixing of two ions under very controlled conditions (Adams *et al*., 2003). Merged beams (Moseley *et al*., 1975) and flowing afterglow (Smith and Adams, 1983) have been employed to study these processes. The construction of the DESIREE double storage ring will allow investigations of mutual neutralization reactions of interstellar anions ($C_4H^-$, $C_6H^-$ and $C_4H^-$) with cations under interstellar conditions, since a very high vacuum and the cooling of the whole ring to temperatures less that 10 K are envisaged for this device (Schmidt *et al.*, 2013).

It would be beyond the scope of this review article to discuss the importance of all gas-phase neutral and ion reactions. Regarding the latter processes, the reader is referred to the review by Larsson et al (2012). However, it can be seen from this summary that a multitude of experimental devices have been used to study gas-phase reactions in the ISM. All these approaches have their advantages and shortcomings, so disagreements in reaction rates and product branching ratios obtained by different methods still exist. However, for many important interstellar chemical gas phase processes, still very little experimental data exist. This is unfortunate, since model calculations need reliable kinetic parameters of these reactions as input data. New devices in the future will hopefully amend this lack of information.

A critical collection of data on different astrophysically relevant gas-phase reactions is provided by the KIDA (Kinetic Database for Astrochemistry) database (Wakelam et al. 2015). It can be accessed via the web under http://kida.obs.u-bordeaux1.fr/.

## 2.2 Experimental investigations of reactions on the surface of interstellar dust grains

It is generally recognized that surface processes on dust grains play an important role in the synthesis of hydrogen molecules and water and organic molecules. Although sequences of gas-phase chemical reactions can lead to the formation of a multitude of molecules, in many cases they are inefficient in interstellar environments because of the presence of rate-limiting processes along the reaction chain. For example, in the gas phase, $H_2$ formation through recombination (i.e., radiative association) of two hydrogen atoms in the ground state is forbidden. Instead, $H_2$ formation requires associative detachment, $H^- + H \rightarrow H_2 + e^-$. Formation of $H^-$ by radiative electron attachment is slow and acts as the rate-limiting process, and thus the latter process was only dominant in the late stages of the primeval universe. Grain-surface processes have several advantages compared with gas-phase reactions as follows: (1) energetic processes like photolysis and ion bombardment of grain ice mantles, which contain many chemical species, can efficiently produce complex molecules; (2) even reactions characterized by small cross sections, which rarely occur through single collisions in the gas phase, may proceed because reactants stay nearby and interact for a long time on the grain surface at very low temperatures; and (3) the surface acts as an absorber of excess energy of reaction, and thus recombination—such as $H_2$ formation and addition reactions—occurs efficiently without dissociation. Formation of hydrogenated molecules, for example, organic molecules, particularly benefits from these advantages on grain surfaces for a number of reasons. First, adsorbates can easily react with hydrogen produced by photolysis or ion bombardment of $H_2O$, which is the major component of ice mantles. Second, with reference to advantage (2), hydrogen addition reactions, even if characterized by moderate activation energies, can be facilitated through quantum tunneling at very low temperatures, where thermally activated reactions are highly suppressed. Quantum tunneling becomes feasible when the de Broglie wavelength of particles is comparable to the width of the activation barrier. In other words, tunneling is particularly effective for reactions involving light atoms, such as hydrogen, at very low temperatures because the wavelength is proportional to $(mT)^{-1/2}$. In addition, radical–radical recombination and tunneling reactions do not require any external energy input, like photons and ions, to proceed. This is particularly crucial for molecular synthesis in dense clouds, where radiation fields are very weak.

Among all hydrogenated molecules ever found in the actual ice mantles of interstellar grains, relatively simple molecules such as $H_2O$, $CH_4$, $NH_3$, $H_2CO$, and $CH_3OH$ can be formed by simple H-atom addition to more primordial species on the grain surfaces. In fact, in recent years many studies have been published about the formation of these molecules through grain-surface reactions of H atoms (for a review, see Watanabe and Kouchi 2008). The formation processes of these molecules compete with each other. For example, methane can be produced by successive hydrogenation of C atom, while methanol results from hydrogenation of CO on grains. The abundance of the simple hydrogenated molecules is strongly related with hydrogenation reaction rates and accretion rates of parent species: C, CO, and N. In fact, it was proposed that the formation of abundant carbon chains observed towards warm protostellar cores can be triggered by sublimation of $CH_4$ molecules from dust grains, where accretion of C atoms on dust exceed that of CO (Sakai and Yamamoto, Chem. Rev. 113, 8981 (2013)). In contrast, in star forming regions where the carbon chain molecules are deficient, the accretion rate of CO would be larger and lead to $H_2CO$ and $CH_3OH$. It is therefore useful to overview the formation processes of not only $CH_4$ but also other competing molecules. In this section, the formation mechanisms of such hydrogenated molecules on cold grain surfaces are briefly reviewed, with a focus on experimental results.

CH4 and NH3

Methane has been observed in both the gas (Boogert *et al.* 1998) and solid (Boogert *et al.* 1996) phase in the direction of high-mass and low-mass protostars. In addition to gas-phase formation, surface reactions are required to explain both the presence of $CH_4$ in polar ice and its low gas/solid abundance ratio. Similarly, both solid and gaseous $NH_3$ have been found at higher abundances than $CH_4$ in a number of different objects. Although a series of ion–molecule reactions have been proposed to form $NH_3$ from N atoms or $N^+$ ions, grain-surface reactions are often considered as a major formation route. The most plausible route for $CH_4$ and $NH_3$ formation on grain surfaces is through successive H-atom addition (hydrogenation) to C and N atoms ($C \rightarrow CH \rightarrow CH_2 \rightarrow CH_3 \rightarrow CH_4$ and $N \rightarrow NH \rightarrow NH_2 \rightarrow NH_3$), respectively. The number of experiments focusing on hydrogenation of C and N atoms on surfaces is still rather limited because of technical difficulties related to C- and N-atom preparation. Hiraoka *et al.* sprayed atomic hydrogen onto C (Hiraoka *et al.* 1998) and N (Hiraoka *et al.* 1995) atoms in CO and $N_2$ matrices, respectively, in the temperature range of 10–30 K. They subsequently detected $CH_4$ and $NH_3$ using temperature-programmed desorption (TPD) methods, although it was not clear whether $CH_4$ and NH$_3$ were formed at 10 K or during the subsequent heating. Recently, Hidaka *et al.* (2011) observed NH$_3$ formation in situ upon the exposure of N atoms in an N$_2$ solid to H atoms at 10 K using Fourier-transform infrared reflection-absorption measurements. Unfortunately, the experimental conditions of these studies were not very relevant for realistic grain surfaces. Use of ice, silicate, or carbonaceous material substrates is desirable to appropriately simulate grain-surface processes. Nevertheless, it is reasonable to infer that the successive hydrogenation of C and N atoms proceeds on the grains because every step is a radical–radical barrierless or low-barrier reaction, assuming a physisorption system. The formation rate of the radical–radical hydrogenation reaction on grain surfaces at very low temperatures should be limited by the surface diffusion of atomic hydrogen or the accretion rates of C or N atoms. Surface diffusion of atomic hydrogen was recently investigated experimentally on amorphous solid water (ASW) at 10 K. From direct measurements of H atoms on ASW, it was found that most adsorption sites on ASW are sufficiently shallow for H atoms to thermally migrate from site to site, even at 10 K (Watanabe *et al.* 2010; Hama *et al.* 2012). Therefore, once H, C, and N atoms land on the grain surface, the diffusion rate of H atoms would not suppress the formation of CH$_4$ and NH$_3$. Instead, exothermic abstraction reactions with H atoms from intermediate radicals, for example, NH + H $\rightarrow$ N + H$_2$ and CH + H $\rightarrow$ C + H$_2$, may compete with addition reactions and slow down the formation of CH$_4$ and NH$_3$. Experiments on this kind of abstraction reaction are highly desirable.

<u>H$_2$CO and CH$_3$OH</u>

Since solid H$_2$CO and CH$_3$OH have been found abundantly in ice mantles, their formation mechanisms have been thoroughly investigated. Theoretical studies first suggested that these molecules are formed through successive hydrogenation of CO on grains, that is, CO $\rightarrow$ HCO $\rightarrow$ H$_2$CO $\rightarrow$ CH$_3$O (or CH$_2$OH) $\rightarrow$ CH$_3$OH (e.g., Tielens and Whittet 1997). There is a critical difference between the formation of H$_2$CO and CH$_3$OH on the one hand and CH$_4$ and NH$_3$ on the other. The successive hydrogenation of CO, leading to H$_2$CO and CH$_3$OH, includes radical-formation processes, CO + H $\rightarrow$ HCO and H$_2$CO + H $\rightarrow$ CH$_3$O, which have significant activation barriers in the approximate energy in temperature range of 2000–3000 K (Woon 2002).

Conversely, successive hydrogenation of C and N atoms consists of radical–radical barrierless or low-barrier reactions only. At very low grain temperatures, $H_2CO$ and $CH_3OH$ formation through thermally activated hydrogenation is highly inhibited and thus requires quantum-tunneling reactions. Because the CO reactant can be readily prepared on a cold substrate in laboratory settings, CO hydrogenation has been targeted by many experimentalists. Since the first attempt where $H_2CO$ formation was observed by using the TPD method (Hiraoka *et al.* 1994), a number of improved experiments have been performed (e.g., Watanabe and Kouchi 2002; Watanabe *et al.* 2004; Fuchs *et al.* 2009). Watanabe and Kouchi (2002) demonstrated experimentally for the first time that $H_2CO$ and $CH_3OH$ are both produced at 10 K by exposure of $H_2O$–CO amorphous ice to cold H atoms. In subsequent experiments, they elucidated the effects of various parameters—such as surface temperature and composition (Watanabe *et al.* 2004, 2006)—on hydrogenation of CO. The same reaction system has been investigated by two other groups, and essentially the same results were obtained (Fuchs *et al.* 2009; Pirim and Krim 2011). The main findings of these experiments were the following: $H_2CO$ and $CH_3OH$ are efficiently formed by CO hydrogenation through quantum tunneling on dust-grain analogues even at 10 K. Upon exposure of CO on ASW to H atoms with fluences equivalent to those in molecular clouds over $10^5$–$10^6$ yr, yields of $H_2CO$ and $CH_3OH$ reproduced abundances of the solids observed in ice mantles (Watanabe *et al.* 2003). To evaluate more quantitatively whether this hydrogenation process works in realistic interstellar conditions, competing reactions should be considered. Among reactions involving hydrogen atoms on grain surfaces, $H_2$ formation through H–H recombination is the most important process. The CO hydrogenation rate is proportional to the surface number density of H atoms, while the recombination rate is proportional to its square. In laboratory experiments, $H_2$ formation must be enhanced because of the expected higher H-atom number density. This can be understood when we realize that the H-atom fluxes used are several orders of magnitude larger than those in interstellar environments. As a result, $H_2CO$ and $CH_3OH$ formation would be rather more efficient in interstellar clouds than in the experiments. The impact of the surface composition and temperature on CO hydrogenation is also an astrochemically important factor. The CO hydrogenation rate on ASW gradually decreases as the temperature increases to 15 K and drops drastically at temperatures greater than 20 K (Watanabe *et al.* 2004). This temperature dependence can be attributed to the significantly reduced residence time of H atoms (i.e., to the number density of H atoms) on the ASW surface at those temperatures. That is, the desorption rate of H atoms exceeds the tunneling-reaction rate with increasing temperature. Experimental results indicate that CO hydrogenation by "nonenergetic" H atoms would be inefficient on grains at temperatures in excess of 20 K. Note that translationally energetic H atoms produced by photolysis or ion bombardment of molecules like $H_2O$ can still react with CO even at temperatures greater than 20 K within ice mantles, because such energetic atoms can overcome the reaction's activation barrier immediately and, therefore, tunneling which requires significant residence time on the surface is not necessary. The role of solid $H_2O$, the main component of ice mantles, in the context of CO hydrogenation has been highlighted experimentally by comparison with results related to the hydrogenation of pure solid CO. The hydrogenation rate of CO on ASW is almost equivalent to that of pure solid CO around 10 K for the same H-atom flux. However, at temperatures above 12 K, the difference in hydrogenation rates between both systems becomes considerable. The CO hydrogenation rate on pure solid CO is approximately three orders of magnitude smaller than that on ASW at 15 K (Watanabe and Kouchi 2008). Since the CO hydrogenation rate is proportional to both the residence time of H atoms on the surface and the tunneling transmission rate for the reaction's potential barrier, the ASW surface should enhance either or both compared with pure solid CO surfaces. However, it is unlikely that the existence of $H_2O$ molecules substantially modifies the shape of the potential barrier for $CO + H \rightarrow HCO$, which governs the tunneling transmission rate, because interactions of both H and CO with $H_2O$ are of van der Waals type. Therefore, enhancement of hydrogenation on ASW would be attributed to an increase in the residence time, that is, an increase of the surface number density of H atoms. It is plausible that a surface composed of molecules with a larger dipole moment produces a higher physisorption energy for H atoms. The dipole moment of $H_2O$ is approximately 20 times larger than that of CO, so that H atoms would stay longer on ASW than on pure solid CO. The effect of the ASW surface on the temperature dependence of the tunneling reactivity varies with the reactions in the physisorption system. Fast reactions, which do not need long interaction times with H atoms, should be less affected by the ASW.

Interstellar formaldehyde (Loinard *et al.* 2000) and methanol (Parise *et al*., 2006) are well-known, highly deuterated species. The high deuterium enrichment in these molecules cannot be explained easily by invoking pure gas-phase chemistry. It is reasonable to consider that surface reactions contribute significantly to the formation of deuterated formaldehyde and methanol, as they do to the formation of normal $H_2CO$ and $CH_3OH$. Theoretical models (Charnley *et al.*, 1997) have proposed successive additions of H and D atoms to CO on grain surfaces as a route for the formation of deuterated formaldehyde and methanol. Hidaka *et al.* (2007) first quantitatively evaluated deuteration of CO on ASW and found that deuterium addition, $CO + D \rightarrow DCO$, is slower by approximately one order of magnitude than hydrogen addition, and that simple addition of D atoms to CO is inefficient. When considering the isotope effect of the tunneling reaction, this is readily understandable. Tunneling reactions are very sensitive to the tunneling mass and become considerably slower for heavier isotope masses. Nagaoka *et al.* (2005) first revealed experimentally that H–D substitution reactions occur on low-temperature surfaces following cold (30–100 K) D-atom exposure of solid $CH_3OH$, that is, $CH_3OH \rightarrow CH_2DOH \rightarrow CHD_2OH \rightarrow CD_3OH$, probably via a sequence of H abstraction and D addition. The backward processes leading to a reduction in the number of D atoms in molecules (e.g., $CH_2DOH \rightarrow CH_3OH$) was not observed upon exposure to H atoms. Isotopologues containing D in the hydroxyl group (Me–OD) have not been observed. Similarly, deuterated formaldehyde is produced very efficiently by H–D substitution of normal formaldehyde (Hidaka *et al.*, 2009). These results have first indicated that H-D substitution reactions on the grain surfaces work for deuterium enrichment of formaldehyde and methanol. The non-detection of Me–OD in the experiments is consistent with the low abundances of Me–OD molecules observed. Based on these experimental results, Taquet *et al.* (2012) recently proposed a model that showed that D- and H-atom abstraction and substitution reactions are crucial in grain-surface chemistry and should be incorporated in models.

### $H_2O$

$H_2O$ is the most abundant component observed in ice mantles. Its abundance cannot be explained merely by gas-phase synthesis (e.g., Hasegawa *et al.*, 1992). The formation of solid $H_2O$ has been intensively studied experimentally in recent years. The simplest route of $H_2O$ formation is by successive addition of H atoms to an oxygen atom ($O \rightarrow OH \rightarrow H_2O$). Hiraoka *et al.* (1998) and Dulieu *et al.* (2010) reported that this formation route works on cryogenic surfaces. In their experiments, nondissociated $O_2$ molecules and $O_3$, which result from the surface reaction $O + O_2$, coexist on the sample surfaces. Since these molecules can be additional sources for $H_2O$, as described below, it is not easy to determine the formation efficiency of $H_2O$ through successive hydrogenation of O atoms alone under such experimental conditions. Nevertheless, it would be safe to consider that successive hydrogenation of O atoms proceeds efficiently on grain surfaces, because every step is radical–radical reaction. A theoretical model has predicted that successive H-atom addition to $O_2$ molecules ($O_2 \rightarrow HO_2 \rightarrow H_2O_2 \rightarrow H_2O + OH$), combined with the reaction $OH + H_2 \rightarrow H_2O + H$, considerably contributes to $H_2O$ formation in dense molecular clouds (Cuppen and Herbst 2007). The final step of the former process, $H_2O_2 + H \rightarrow H_2O + OH$, and the latter reaction, $OH + H_2 \rightarrow H_2O + H$, have significant barriers and thus rarely occur at grain temperatures through thermally activated reactions. Miyauchi *et al.* (2008) demonstrated experimentally that successive hydrogenation of $O_2$, leading to the formation of $H_2O$, occurs efficiently even at 10 K through quantum tunneling. Subsequently, Ioppolo *et al.* (2008) also confirmed the efficacy of this process. Recently, a large isotope effect on the reaction $H_2O_2 + H \rightarrow H_2O + OH$ has been reported as a feature of the tunneling reaction (Oba *et al.*, 2014). For the reaction of OH with $H_2$, Oba *et al*. (2012) presented the first experimental evidence of solid $H_2O$ formation through the tunneling reaction of OH (in its electronic and vibrational ground states) + $H_2$ at 10 K. Using isotopologues such as OD and $D_2$, they found that a significant isotope effect results from differences in the effective mass of each reaction. $H_2O$ can also be formed through hydrogenation of $O_3$ (Mokrane *et al*. 2009; Romanzin *et al*. 2011). However, $O_3$ has not been observed in the interstellar medium, in neither gas nor solid phases. In fact, Taquet *et al*. (2013) predicted that $O_3$ is a minor product on grain surfaces.

## 2.3   Experiments simulating reactions in ice mantles of interstellar grains

Reactions of $CH_4$ / $CH_3OH$ in simulated ice mantles of interstellar grains

Solid $CH_4$ has been observed towards low- and high- mass young stellar objects (Lacy *et al.*, 1991; Boogert *et al*., 1996, 1997; Gibb *et al*., 2004; Oberg *et al*.,2008). Its abundance with respect to solid water varies in the range 2-10% (Boogert *et al.*, 1997; Oberg *et al.*, 2008). $CH_4$ is believed to be formed on grains either by hydrogenation (Aikawa *et al.*, 2005; Oberg *et al.*, 2008) of accreted C-atoms or by energetic processing of $CH_3OH$-rich ices (Boogert *et al.*, 1996; Baratta *et al.*, 2002; Garozzo *et al.*, 2011). In fact, it is widely accepted that icy grain mantles are continuously processed by low-energy cosmic rays and UV photons (Jenniskens *et al.*, 1993; Shen *et al.*, 2004). Shen *et al.* (2004) estimated the energy deposition onto water-ice grain mantles by cosmic rays and by UV photons in dense molecular clouds. They found that depending on the assumed cosmic-ray spectrum at low energy, after $10^7$ years the dose deposited by UV photons varies in the range 100-10 eV/molecule and the dose deposited by cosmic rays varies in the range 10-1 eV/molecule.

In the past 35 years, several experiments have been performed to study the effect of ion bombardment and UV photolysis on the chemical composition of icy samples. In fact, several laboratories worldwide are involved in this kind of study (e.g., Sandford *et al.,* 1988; Palumbo and Strazzulla 1993; Gerakines *et al.,* 1996; Cottin *et al.,* 2003; Bennett *et al.,* 2006; Oberg *et al.,* 2009; Fulvio *et al.*, 2009; Boduch *et al.,* 2012; Modica *et al.,* 2014; Munoz Caro *et al.*, 2014). Icy samples are prepared in a high vacuum (HV; P $\sim10^{-7}$ mbar) chamber or ultra high vacuum (UHV; P $\sim10^{-9}$ mbar) chamber. An infrared-transparent substrate (e.g., crystalline silicon or KBr) is placed in thermal contact with a cryostat and cooled down to 10-20 K. Gases are admitted into the vacuum chamber in order to accrete a thin ice film (0.1-10 µm). Icy samples can be processed by fast ions (E= keV-MeV) or UV photons (obtained by a discharge lamp or synchrotron radiation). The ice samples can be analyzed before and after processing by infrared transmission spectroscopy or Refection Absorption IR (RAIR) spectroscopy.

Laboratory experiments have shown that $CH_4$ is easily destroyed by ion bombardment or UV photolysis (e.g., Gerakines *et al.*, 1996; Baratta *et al.*, 2002). As an example, Figure 5 shows the column density of $CH_4$ after ion bombardment of pure $CH_4$ and a mixture $H_2O$: $CH_4$=4:1 at 12 K. Column density values are plotted versus dose (given in eV/16u). The dose is obtained from the knowledge of the ion fluence impinging on the sample (ions $cm^{-2}$) measured during the experiment and the stopping power (eV $cm^2$ $molecule^{-1}$) obtained using SRIM 2008 code (Ziegler *et al.*, 2008). As suggested by Strazzulla and Johnson (1991), the dose given in units of eV per small molecule (16u) is a convenient way to characterize chemical changes and compare the results obtained after processing of different samples. Experimental data reported in Fig. 5 are adapted from Baratta *et al.* (2002) and Garozzo *et al.* (2011). The column density is obtained from the $CH_4$ band at about 1300

cm$^{-1}$ using a band strength value equal to $6.4 \times 10^{-18}$ cm molecule$^{-1}$ (Mulas *et al*., 1998) after the transmission spectra are converted to optical depth scale. The column density of CH$_4$ at a given dose is divided by the value measured soon after deposition (i.e., before bombardment starts). We notice that the CH$_4$ column density ratio has values greater than 1 at low dose. As pointed out by Garozzo *et al*. (2011), this is due to the variation in the band strength value after ion bombardment as also observed for other bands in other molecules (e.g., Leto and Baratta, 2003; Loeffler *et al*., 2005).

FIG. 5.  Normalized column density of CH$_4$ as a function of irradiation dose (eV/16u) after ion bombardment of pure CH$_4$ and a H$_2$O:CH$_4$=4:1 mixture at 12 K with 30 keV He$^+$.

After bombardment, new absorption bands are observed in the infrared spectra, which indicate the formation of additional volatile species such as ethane (C$_2$H$_6$), propane (C$_3$H$_8$), ethylene (C$_2$H$_4$), and acetylene (C$_2$H$_2$). When CH$_4$ is mixed with H$_2$O and/or N$_2$, other and more complex species are formed (e.g., Moore and Hudson, 1998, 2003; Baratta *et al*., 2002, 2003). In any case, after further bombardment a refractory residue is formed as demonstrated by the appearance of the amorphous carbon feature in the Raman spectra (e.g., Ferini *et al*., 2004; Palumbo *et al*., 2004). Jones and Kaiser (2013) studied the effects of electron irradiation of pure CH$_4$ by reflectron time-of-flight mass spectrometry and found that high-molecular-weight hydrocarbons of up to C$_{22}$, among them alkanes, alkenes, and alkynes, are formed; Paardekooper et al. (2014) and Bossa et al. (2015) investigated the effect of UV photolysis on pure CH$_4$ ice at 20 K combining laser desorption and time-of-flight mass spectrometry showing the formation of large C-bearing species.

Observations have shown that in icy grain mantles the abundance of methanol (CH$_3$OH) with respect to H$_2$O along the line of sight of high-mass and low-mass young stellar objects spans over a large range: 1%-30%   (Gibb *et al*., 2004; Boogert *et al*., 2008).

Recent laboratory experiments (e.g., Hiraoka *et al*., 2002; Watanabe and Kouchi, 2002; Fuchs *et al*., 2009) have shown that CH$_3$OH molecules efficiently form after hydrogenation of CO molecules in CO-rich and water-poor ices. Therefore, it is reasonable to assume that CH$_3$OH-rich and water-poor ice mantles may exist along the line of sight of dense molecular clouds (e.g., Skinner *et al*., 1992; Palumbo and Strazzulla, 1992; Cuppen *et al*., 2011). When methanol is processed by energetic ions, the column density of pristine methanol decreases and new bands appear in the infrared spectra indicating the formation of other, also complex, species (e.g., Palumbo *et al.*, 1999; Baratta *et al.*, 2002; Oberg *et al.*, 2009; Bennett *et al.*, 2007).    Figure 6 (top panel) shows the normalized column density of methanol after ion bombardment at 16 K of pure CH$_3$OH and CO:CH$_3$OH and N$_2$:CH$_3$OH    mixtures as a function of dose in eV/16u. Experimental data are adapted from the works of Modica and Palumbo (2010) and Islam *et al*. (2014). The column density of methanol is obtained from the band at about 1030 cm$^{-1}$ using the band strength value of $1.3 \times 10^{-17}$ cm molecule$^{-1}$ (Palumbo *et al*., 1999). Figure 6 (bottom panel) shows the ratio between the column density of CH$_4$ formed after processing and the column density of CH$_3$OH at the same dose. We notice that this ratio is independent of the initial mixture within experimental uncertainties. Several other species are formed after ion bombardment of CH$_3$OH-rich mixtures such as carbon monoxide (CO), carbon dioxide (CO$_2$), formyl radical (HCO), formaldehyde (H$_2$CO), ethylene glycol (C$_2$H$_4$(OH)$_2$), methyl formate (HCOOCH$_3$), and glycolaldehyde (HCOCH$_2$OH) (e.g. Moore *et al*., 1996; Palumbo *et al*., 1999; Hudson and Moore, 2000; Modica and Palumbo. 2010).

FIG. 6. Top panel: Normalized column density of CH$_3$OH as a function of irradiation dose (eV/16u) after ion bombardment of pure CH$_3$OH and two mixtures, CO:CH$_3$OH and N$_2$:CH$_3$OH, at 16 K with 200 keV H$^+$. Bottom panel: Column density of CH$_4$ formed at 16 K after ion bombardment (200 keV H$^+$) of the same CH$_3$OH-rich ice mixtures divided by the column density of CH$_3$OH at the same dose.

The experimental results described here could be added to chemical models to understand the equilibrium value reached between CH$_4$ formation processes (i.e., hydrogenation of C-atoms and energetic processing of $CH_3OH$-rich ices) and the destruction processes induced by ion bombardment.

Ten different molecular species have been firmly identified in interstellar icy grain mantles. Among them $CH_4$ and $CH_3OH$ have been detected towards both high-mass and low-mass young stellar object. From the data reported by Gibb *et al.* (2004), it is possible to estimate that, towards high-mass young stellar objects, the column density ratio $CH_4/CH_3OH$ is on average less than one. On the other hand, this ratio is higher, as high as two, towards low-mass young stellar objects (Oberg *et al.*, 2008; Boogert *et al.*, 2008). This difference could be ascribed to the different average ice temperature in the line of sight to high- and low-mass objects along with the different sublimation temperature of pristine $CH_4$ (about 30 K) and pristine $CH_3OH$ (about 130 K; e.g., Collings *et al.*, 2004). However, it is important to note that volatile species can remain trapped in the refractory residue formed after ion bombardment and can be observed in the spectra taken at temperatures higher than their sublimation temperature (e.g., Palumbo *et al.*, 1999; Sicilia *et al.*, 2012).

As shown in Fig. 6 (bottom panel), laboratory experiments show that the column density ratio $CH_4/CH_3OH$ after ion bombardment of $CH_3OH$-rich ice is less than one in the dose range investigated   (Palumbo et al., unpublished data). This suggests that ion bombardment of $CH_3OH$-rich ices can be relevant but cannot be the only formation route to observed solid $CH_4$; hydrogenation of accreted carbon atoms is a relevant process too.

It is generally accepted that ice grain mantles are processed simultaneously by UV photons and low-energy cosmic rays. Several investigations on the comparison between the two processes have been carried out (e.g., Gerakines *et al.*, 2000, 2001; Cottin *et al.*, 2001; Baratta *et al.*, 2002; Munoz Caro *et al.*, 2014), and recently investigations on the effects obtained after simultaneous processing have been performed (Islam *et al.*, 2014). Experimental results have shown that, from a qualitative point of view, ion bombardment and UV photolysis generate similar changes in interstellar ice analogs. However, quantitative differences between the two processes have been observed.

Even if only ten species have been firmly identified, it is generally accepted that other, more complex molecules are also present in icy grain mantles which cannot be detected in the solid phase by IR spectroscopy. These species are expected to enrich the gas phase composition after desorption of icy grain mantles (e.g., Palumbo *et al.*, 2008: Modica and Palumbo, 2010) and could be incorporated in planetesimals and comets. To check the role of ion bombardment in the formation of complex species after ion bombardment of $CH_4$ and $CH_3OH$-rich ices, we have compared the profile of the $CH_4$ band at about 1300 cm$^{-1}$ observed towards dense molecular clouds with the profile we obtain in our laboratory spectra after ion bombardment. As an example, Fig. 7 shows the comparison between the band profile observed with the ISO satellite towards the high-mass young stellar object NGC7538 IRS9 and the profile of the $CH_4$ band in different laboratory spectra. Figure 7 shows that a good comparison is obtained considering the contribution of two laboratory spectra, namely, a mixture $H_2O:CH_4$ after ion bombardment at low temperature and a $CH_3OH$-rich ice after ion bombardment at low temperature and warm up to 100-125 K. In each panel, the red line is the best-fit to the observed data points obtained by a linear combination of two laboratory spectra. The thin lines are the laboratory spectra scaled by the coefficient given by the fit procedure. This result is consistent with the hypothesis that $CH_4$ formed after C hydrogenation is present on cold ice grain mantles along the line of sight and $CH_4$ formed after energetic processing of $CH_3OH$ is also there. Here, the temperature at which laboratory spectra are taken has to be regarded as the value at which the profile of the $CH_4$ band better reproduce the average profile along the line of sight due to icy grain mantles at different temperatures.

FIG. 7. Comparison between observed and laboratory spectra.

Many recent results support experimental efforts (e.g., Allodi *et al.*, 2013) to use more sensitive techniques to clearly show the formation of complex molecules and/or fragments that could be of primary relevance for astrobiology and an understanding of which species should be searched for, by ground-based or space-born facilities, in protostellar environments, protoplanetary disks, and atmospheres of extrasolar planets or moons that afford evidence for the presence of a complex chemistry that could evolve toward a biosphere.

Formation of Complex Organic Compounds in Simulated Ice Mantles of Interstellar Grains

It was shown that interstellar dust particles in dense clouds are covered with ice mantles that are composed of water, carbon monoxide, carbon dioxide, methanol, methane, and ammonia.   The question is whether more complex organic compounds could be synthesized in the ice mantles.   Greenberg proposed the following scenario: (1) Complex organic molecules were formed in ice mantles of interstellar dust particles (ISDs) in dense clouds: (2) The ISD complex molecules were altered by photochemical and thermal processes in diffuse cloud, protosolar nebulae, etc., (3) The ISDs with organic molecules are then aggregated to form comets (Greenberg and Li, 1997). It is not possible to observe complex organic compounds in the ice mantles; the only way to examine this is in laboratory simulation experiments.

The major energy for the reactions in ice mantles seems to be cosmic rays and cosmic ray-induced ultraviolet light since stellar ultraviolet light cannot penetrate deeply into dense clouds. To simulate possible reactions in ice mantles of interstellar dust particles, frozen mixtures of simulated interstellar media were irradiated with ultraviolet light or high energy protons.

Greenberg and coworkers pioneered a laboratory simulation of photochemical formation of complex organic molecules from possible interstellar media.   In their early experiments, a frozen mixture of $H_2O$, $NH_3$, $CH_4$, and/or CO was used as simulated ice mantles of ISDs.   The starting gas mixtures were frozen onto aluminum substrates that had been cooled down to 12 K, and they were irradiated in vacuum with UV light from a hydrogen lamp. The products were warmed up to room temperature, and the residues on the substrates were analyzed by GC/MS after trimethylsilyl (TMS)-derivatization. When a mixture of $H_2O$, $NH_3$, and CO was used, various molecules including glycine were identified in the products. When $CH_4$ was used in place of CO, no residues were observed (Briggs et al., 1992).   Thus, several photolysis experiments after this research tended to use CO and/or $CH_3OH$ rather than $CH_4$.

Muñoz Caro *et al.* (2002) irradiated a frozen mixture of $H_2O$, CO, $CO_2$, $CH_3OH$, and $NH_3$ in a vacuum with UV light from a hydrogen lamp at 12 K.   Bernstein et al. (2002) also used a frozen mixture of $H_2O$, CO, $CH_3OH$, HCN, and $NH_3$ as a target of VUV irradiation at 15 K.   Both teams detected a number of racemic amino acids after hydrolysis of the products.

To examine direct actions of cosmic ray particles, Kobayashi *et al.* irradiated a frozen mixture of $H_2O$, $CH_4$ (or CO), and $NH_3$ with 3 MeV protons from a van de Graaff accelerator and confirmed the formation of amino acid precursors (compounds giving amino acids after hydrolysis) (Kobayashi *et al.*, 1995, Kasamatsu *et al.*, 1997a, b). In particles irradiation experiments, not only CO but also $CH_4$ gave amino acid precursors.

The results suggest that complex molecules including amino acid precursors could be formed in ice mantles of interstellar grains containing $H_2O$, $CH_4$ (or CO, $CH_3OH$), and $NH_3$ by the action of cosmic rays and cosmic ray-induced UV.   In addition, thermal energy during warming of the ice mantles from dense cloud environments to warmer environments would lead to the formation of more complex organics (Theule et al., 2013).   Such organic compounds could have been incorporated into small bodies such as asteroids and comets after a solar system was formed from the materials in dense clouds via a solar nebula. Furthermore, it is of interest to study how interstellar complex organic compounds were altered to cometary and meteoritic complex organic compounds by photochemical, radiochemical, and thermal / hydrothermal / hydrolytic alterations.

## 3.   Experiments simulating submarine hydrothermal systems

Submarine hydrothermal vents were first discovered at the Galapagos spreading center in the late 1970s (Corliss *et al.*, 1979).   There discovery along the boundaries of tectonic plates were regarded as favorable sites for chemical evolution toward the generation of life on Earth for the following reasons:

1) Reducing environments are favorable for abiotic synthesis of organic compounds of biological interest such as amino acids. Though it is controversial whether methane and other hydrocarbons were present in the primitive Earth atmosphere, submarine hydrothermal systems are promising sites where methane is available, because methane is present in hydrothermal systems even in the present oxidizing Earth environments.

2) Thermal energy from magma could provide an energy source for chemical evolution. Though long-term heating at high temperature would destroy abiotically formed organic compounds, those formed in hot environments could be thermal quenched when they erupted into cold seawater.

3) The interaction between hydrothermal fluid and rocks provided high concentration of various metal ions such as manganese, iron, and zinc, which could serve as catalysts for abiotic synthesis of organic compounds (Kobayashi and Ponnamperuma, 1985).

4) Phylogenetic relationships between microorganisms, when displayed on the 16S rRNA universal tree, suggest that the last universal common ancestor of terrestrial organisms (LUCA) was a thermophile, which implies that the earliest terrestrial microbes inhabited hot environment such as the vicinity of submarine hydrothermal vents.

A wide variety of experiments have been performed to simulate possible reactions in submarine hydrothermal systems (SHSs) to study the origins of methane in SHSs and formation and stability of organic compounds, such as amino acids in SHSs. Holm and Andersson (2005) published eleven years ago in Astrobiology a review of hydrothermal organic chemistry experiments carried out up to that date. Recent development of experimental systems simulating submarine hydrothermal systems is summarized in Suzuki et al. (2015).

## 3.1 Experiments simulating methane formation in hydrothermal systems

Not many experiments have been designed to specifically monitor the formation of methane in geothermal systems. Most systems have been designed to monitor as many organic compounds as possible. French (1962, 1970) was the first scientist to recognize hydrothermal systems as sites for abiotic synthesis of organic compounds. He suggested, on the basis of experimental results, that organic compounds are formed abiotically by processes like Fischer-Tropsch type (FTT) reactions between a high pressure gas phase and siderite or other carbonates plus iron-bearing minerals. He also proposed that suitable conditions could be attained by hydrothermal activity at moderate depth within Earth's crust, where the existence of a $CH_4$-rich phase would be favored by low fugacity of oxygen ($fO_2$). Simulations of organic geochemistry in hydrothermal system conditions following French have been entirely focused on more complex organic compounds other than methane, particularly hydrocarbons and amino acids.

About the only hydrothermal simulation experiments that have been designed to focus on the formation of $CH_4$ (together with $H_2$) were performed by Neubeck *et al.* (2011, 2014). They conducted a series of low temperature (30, 50, and 70°C) experiments in which they tested the $CH_4$ and $H_2$ formation potential of olivine. Their results show that not only was hydrogen formed from water because of partial oxidation of Fe(II) in the Fe-silicate fayalite of the olivine, but also that $CH_4$ was produced at these relatively low temperatures from $CO_2$ and $HCO_3^-$. The authors concluded that the presence of spinels like magnetite (formed by the oxidation of the fayalite) and chromite catalyzed the formation of $CH_4$.

### 3.2 Formation of amino acids in SHSs-simulating environments

To examine possible formation of amino acids in submarine hydrothermal systems, Yanagawa and Kobayashi (1992) performed simulation experiments by using an autoclave. Modified hydrothermal vent media (MHVM) containing $Fe^{2+}$ (2 mM), $Mn^{2+}$ (0.6 mM), $Zn^{2+}$ (0.1 mM), $Ca^{2+}$ (0.1 mM), $Cu^{2+}$ (20 mM), $Ba^{2+}$ (0.1 mM), and $NH_4^+$ (50 mM) were prepared, and the pH was adjusted to 3.6.   MHVM in a Pyrex glass tube and a 1:1 mixture of methane and nitrogen (80 kg $cm^{-2}$) were introduced into the autoclave.   They were heated at 325°C for 1.5 – 12 h and then cooled down to room temperature.   The resulting solution was subjected to amino acid analysis by HPLC and GC/MS after acid hydrolysis.

Amino acids such as glycine, alanine, aspartic acid, and serine were detected in the hydrolysates, together with non-protein amino acids like α- and γ- aminobutyric acid, β-alanine, and sarcosine.   The amino acids detected by GC/MS with a chiral column were racemic mixtures.

Procedural blank showed no amino acids.    Thus, it was concluded that amino acid precursors were formed abiotically from methane-containing media in simulated submarine hydrothermal systems (Yanagawa and Kobayashi, 1992).

A major outstanding challenge is quantification of abiotically formed $CH_4$ and other organic compounds in hydrothermal environments on Earth today and on other celestial bodies, as well as improve our understanding of such a process on the early Earth.

## 4. Experiments on methane chemistry in planetary and satellite atmospheres

A wide variety of organic compounds have been detected in extraterrestrial environments. Since there are few relics of chemical evolution on present-day Earth, the inventories of extraterrestrial organic compounds could suggest possible chemical evolutionary pathways toward the generation of life on Earth and elsewhere. Among all the objects in the solar system, Titan and its methane-rich atmosphere provide a natural laboratory of chemical evolution. Thermal escape, and methane reactivity, including its direct photolysis, are responsible for a short lifetime of methane in Titan's atmosphere of about 20-30 Myr (Atreya 2006, Tucker and Johnson 2009, Krasnopolsky 2014). Methane chemistry is, moreover, strongly coupled with nitrogen, leading to heavy hydrocarbon and nitrogen-bearing organic molecules in the gas phase and ultimately to solid organic aerosols that settle downward to Titan's surface. This complex chemistry has been investigated by the Cassini-Huygens space mission and in laboratory experiments. The thickness of Titan's atmosphere leads to two main, isolated reactive-layers in Titan's atmosphere: a layer above the tropopause driven by VUV photochemistry, and a layer below the tropopause driven by soft UV and interactions with the surface.

## 4.1 Reactions in the higher layers of Titan's atmosphere and ionosphere

One of the most surprising discoveries of the Cassini-Huygens mission to the Saturn system was the rich ion-neutral organic chemistry present in the thermosphere and ionosphere of Titan above an altitude of 950 km (Waite *et al.*, 2006). Three observation types characterize that discovery as follows (see Fig. 8):

1) A strong correspondence between the neutral and ion species that were measured below 100 u by the Cassini Ion Neutral Mass Spectrometer (INMS) (see panel A).

2) An extension of the ion mass spectrum up to masses of >250 u measured by the Cassini Plasma Spectrometer (CAPS) Ion Beam Spectrometer (IBS, Crary *et al*., 2009) (see panel B).

3) Large negative ions with masses greater than 5000 u measured by the CAPS Electron Spectrometer (ELS; panel C) that increase in complexity at altitudes below 1000 km (Coates *et al*., 2009).

Westlake *et al*. (2014) demonstrated with correlation analysis that the molecular growth between ions and neutral species below 200 u appears to take place in large part due to addition reactions of $C_2$ compounds ($C_2H_2$ and $C_2H_4$). Westlake *et al*. (2014), Waite *et al*. (2006), and Lavvas *et al*. (2013) argued that the processes of positive and negative ion growth of organic molecules can account for the bulk of the organic synthesis at Titan with largely agglomeration and condensation chemistry occurring below 800 km. This stands in strong contrast to the pre-Cassini-Huygens paradigm of organic synthesis in Titan's stratosphere.

## 4.2 Experiments simulating the higher layers of Titan's atmosphere

The triggering process for methane chemistry in Titan's higher atmosphere is its photolysis. Well characterized at Lyman-α (Romanzin et al. 2008, Gans, Boye et al. 2011), the branching ratios among the photolytic products remain poorly documented out of this wavelength. No measurements exist between the branching ratios for $CH_2$ and $CH_3$ at wavelengths larger than Lyman-α (Gans et al. 2013). Titan's atmospheric models suffer from this lack of experimental data since the 130-140 nm wavelength range dominates methane photolysis in the stratosphere.

From Cassini observations, we know that aerosols not only are bathed in a neutral and ion gas mixture that constitutes Titan's ionosphere, but they also affect the content and charge balance of the gas phase itself (Lavvas et al. 2013). This feedback between ions, neutrals, and solid aerosols is, to date, difficult to predict, but has been successfully approached in the laboratory by dusty-plasma experiments (Coll et al. 1999, Imanaka et al. 2004, Szopa et al. 2006, Trainer et al.

2006, Cable et al. 2012). In this case, partial ionization of the initial $CH_4$-$N_2$ gas mixture is created by a plasma discharge, which is similar to the combined effect of the magnetospheric electrons from Saturn and the VUV solar photons. Plasma experiments have provided support for the interpretation of Cassini-Huygens data. For instance, specific investigation of the ion and neutral gas-phase content have been led by Gautier et al. (2011), Carrasco et al. (2012) and Horvath et al. (2010), involving nitrile molecules, positive ions, and negative ions respectively. Plasma experiments in the laboratory have also demonstrated that the addition of CO to the mixture (perhaps due to oxygen from Enceladus) can result in efficient oxygen incorporation in the aerosols (Fleury et al. 2014) and in amino acid formation (Hörst et al. 2012).

A new type of experiment that mimics Titan's ionospheric chemistry appeared with the development of photoreactors coupled with VUV synchrotron beamlines (Imanaka and Smith 2010, Peng et al. 2013). In the latter case, the dissociation and ionization processes occur through direct VUV photolysis of methane and molecular nitrogen. Neutral chemistry only, as in Titan's stratosphere, is also often simulated with experimental setups that ensure methane dissociation by softer VUV (such as Lyman-$\alpha$ wavelength) and UV sources (Tran et al. 2003, Trainer et al. 2012).

Methane concentration varies significantly in the ionosphere of Titan according to altitude (Waite Jr et al. 2005, Hébrard et al. 2006) from 2 up to about 10%. This parameter is therefore often considered in laboratory simulations of Titan's ionosphere. A strong influence of the methane concentration on the aerosol production efficiency was found by Trainer et al. (2006) and Sciamma-O'Brien et al. (2010), with mass production rates varying by an order of magnitude according to the initial methane concentration. The maximum is reached at intermediate methane concentrations and is driven by methane dissociation. Methane dissociation involves two opposing effects on the aerosol chemical growth: a carbon organic supply and an increase in the release of H atoms in the reactive medium (Carrasco et al. 2012). The methane concentration in the reactive mixture also impacts the final chemical composition of the laboratory aerosols (Derenne et al. 2012, Gautier et al. 2014, He and Smith 2014) and their optical properties (Quirico et al. 2008, Mahjoub et al. 2012, Brassé et al. 2015), suggesting an altitude-dependency of the aerosols in higher layers of Titan's atmosphere.

One of the main experimental limitations of plasma and photolytic reactors is to reproduce Titan's high-altitude pressure conditions (lower than $10^{-4}$ mbar, see Figure 9) and long mean free paths for ions and neutrals in the laboratory. Indeed, at these low pressures the mean free path of the reactant species has to be several orders of magnitudes lower than the smallest dimension of the reactor to prevent prominent wall effects (Carrasco et al. 2013). This condition involves reactors with kilometric dimensions to simulate Titan's gas-phase ionospheric chemistry, which is inaccessible in conventional laboratories. The higher pressures used in the plasma and photolysis setups suggest that the production of aerosols in laboratory experiments can be rather distinct from Titan ionospheric chemistry, as three body-reactions will contribute significantly to the whole chemistry network. Lower pressure plasma experiments under conditions where radicals have marginal lifetimes have been attempted. However, those conditions (pressures in the mbar range with mean free paths of millimeters) differ greatly from the ones encountered in Titan's ionosphere, in which mean free path lengths have the dimensions of kilometers. Thissen *et al.* (2009) attempted to produce Titan chemistry at low pressures using a synchrotron source and a low pressure vessel and found some of the same reaction pathways as are observed at Titan. However, an important contribution of wall-effects was observed and the nitrogen containing hydrocarbon production has remained elusive. To date, no laboratory experiments have been successfully performed that replicate the pressure conditions at Titan where the complex hydrocarbons are produced.

Reactions of interest for Titan's higher atmosphere are moreover individually investigated through single collision experiments, which feed Titan's photochemical models with a bottom-up approach. The experimental setups are, in this case, the same as previously described for the ISM purpose (see section 2.1). A specific focus on Titan's negative ion chemistry has moreover led to an intermediate pressure regime approach in the laboratory. The progressive increase of pressure in an ion-neutral reaction cell enabled workers to probe various multiple collision regimes and generate secondary products for the reaction between $CN^-$ and $HC_3N$ (Žabka et al. 2012).

## 4.3 Reactions in the lower Titan atmosphere

The lower atmosphere of Titan had not been observed before the landing of Huygens probe in 2005 due to dense mist; this is another possible site for abiotic synthesis of organic compounds along with Titan's higher atmosphere.    Solar UV light and electrons from the magnetosphere of Jupiter are two of the major energy sources in the higher Titan atmosphere, but they cannot reach the troposphere of Titan.    Thus, possible energy sources for chemical reactions in the troposphere would be cosmic rays and meteor impacts.

Taniuchi *et al.* (2013) examined the possible formation of organic compounds in a simulated Titan atmosphere by proton irradiation.    A mixture of 5 % methane and 95 % nitrogen (700 Torr) was irradiated with 3 MeV protons from a van de Graaff accelerator (TIT, Japan).    As soon as the irradiation started, mist formed in gaseous phase, showing that solid materials were formed from methane and nitrogen.    They can be called *Titan tholins*, though their appearance is different from *tholins* produced by plasma discharges in a low-pressure mixture of methane and nitrogen.    Hereafter, the solid product by proton irradiation is called *PI Titan tholins*.

PI Titan tholins have quite complex structures as suggested by FT-IR and pyrolysis GC/MS, whose average molecular weight was some hundred Dalton as estimated by gel permeation chromatography.    With pyrolysis, under similar conditions, as performed by the Aerosol Collector and Pyrolyzer (ACP) on board Huygens Probe, $NH_3$ and HCN were observed as the major pyrolysis products, which was the same as the analytical results of Titan aerosol by Huygens ACP.

These could be partly dissolved with water and some organic solvents such as tetrahydrofuran (THF).    When PI Titan tholins were acid-hydrolyzed, a wide variety of amino acids were detected by HPLC.    Glycine was predominant whose energy yield (G-value) was as high as 0.03 molecule/100 eV deposited.    Such chiral amino acids as alanine, a-aminobutyric acids, valine, norvaline, and aspartic acids were also detected by HPLC and proved to be racemic mixtures.    Thus, it was confirmed that the PI Titan tholins included amino acid precursors.

The present starting materials are only methane and nitrogen, which did not contain any oxygen-bearing molecules.    Thus, free amino acids that contain oxygen atoms could not be formed by the irradiation.    To examine the source of oxygen in amino acids obtained here, the PI Titan tholins were hydrolyzed with 1 M HCl prepared by dilution of concentrated HCl with $H_2^{18}O$, and were subjected to MALDI-MS analysis. The results that yielded amino acids mainly contained $^{18}O$, but not $^{16}O$, and showed that oxygen atoms were incorporated during hydrolysis of amino acid precursors.

Short-wavelength UV radiation cannot reach the lower atmosphere of Titan, but near UV ($\lambda > 300$ nm) could be a possible energy source of chemical reaction in the troposphere of Titan. Such UV photons cannot dissociate $N_2$ or $CH_4$, but some photochemical products in stratosphere could fall down to the troposphere to take part in some photochemical reactions. $C_4N_2$ (dicyanoacethylene) was chosen as a candidate molecule that could be supplied from the stratosphere to the troposphere of Titan after photolysis of $C_4N_2$ yielded non-volatile, haze-like materials (Gudipati et al., 2012; Couturier-Tamburelli et al., 2014).

Thus, it was concluded that tholins could be formed not only in the Titan stratosphere but also in the Titan troposphere. There, tholins could be formed directly from $CH_4$ by cosmic rays and from such activated molecules as $C_4N_2$ (formed from $CH_4$ in stratosphere) even by near UV light.

Titan tholins synthesized under upper atmosphere conditions by plasma discharge also yielded amino acids after acid hydrolysis (Khare *et al.* 1986). Nguyen et al. (2008) also reported that "discharge" tholins gave amino acids and carboxylic acids after acid hydrolysis. It is unlikely, however, that acid hydrolysis with strong acid occurred in the Titan environment. Interaction of tholins (discharge tholins) with possible solvents in Titan environments (ammonia water and hydrocarbons) has been studied. Poch et al. (2012) reported that amino acids and nucleic acid bases were formed when tholins, which were synthesized by glow discharge in a mixture of $N_2$ and $CH_4$ (98:2), were dissolved in ammonia water. Neish et al. (2010) synthesized tholins by glow discharge in a mixture of $N_2$ and $CH_4$ (98:2): They found that a wide variety of amino acids and all five nucleic acid bases were formed when the tholins were dissolve in cold ammonia water (253 or 293 K). To avoid the possibility of contamination, experiments in which isotope-labeled starting materials are used would be favorable.

The energy flux of solar UV and electrons from the Jupiter magnetosphere is much higher than that of cosmic rays. However, amino acid precursors formed in Titan's lower atmosphere by cosmic rays can not be ignored since the energy yield (G-value) of amino acid precursors by proton irradiation is much higher than that of plasma discharge or UV irradiation. Both tholins formed in upper Titan's atmosphere and tholins formed in the lower Titan atmosphere would be supplied to Titan's surface, and could interact with surface water ice or ammonia water from subsurface ocean to yield amino acids. It would be of interest to discriminate upper Titan tholins from lower Titan tholins in future Titan missions.

Besides traditional laboratory simulation experiments, a new trend in simulation studies involves space experiments. By using man-made satellites, space shuttles, and space station facilities, a wide variety of experiments including astrobiology experiments have been conducted (Horneck et al., 2010). Carrasco et al (2015) placed a Titan-type gas mixture (150 kPa) of $N_2$, $CH_4$, and He (or $CO_2$) in cells with $MgF_2$ windows on an exposed facility of the International Space Station. The gas mixture was exposed to solar vacuum UV. Unsaturated hydrocarbons were detected, while conventional laboratory VUV irradiation experiments using low-pressure gas mixtures mainly yielded unsaturated hydrocarbons: The difference in gaseous pressure brought about these results. It is of interest to repeat such space experiments to discern synergetic effects of solar UV and cosmic rays in chemical evolution.

Tholins formed in Titan's atmosphere containing methane would be partly dissolved in the liquid ethane-methane lakes found on Titan (Brown et al., 2008). We have quite limited knowledge as to how tholins would behave when dissolved in cold hydrocarbon lakes. It will be of great interest to simulate possible chemical evolution of tholins in the liquid methane-ethane lakes of Titan.

## 4.4   Origins of methane in the outer solar system

The number of experiments dedicated to the investigation of methane chemistry have been scarce with regard to other planetary environments, most experimental efforts having been dedicated to the investigation of methane formation conditions via Fischer-Tropsch-Type (FTT) reactions. For example, to investigate the origin of Titan's atmospheric methane, Sekine *et al*. (2005) conducted FTT experiments in low pressure and low temperature ranges corresponding to the thermodynamic conditions hypothesized to have occurred in the protosolar nebula and in circumplanetary disks such as Saturn's subnebula. These authors thus showed that the FTT catalysis is efficient in a narrow temperature range (~500-600 K) in the gas phase conditions of the protosolar nebula or Saturn's subnebula and can be used to trace back the evolution of CO and $CO_2$ in these environments. The experimental results of Sekine *et al.* (2005) suggest that these two species are converted into $CH_4$ within time significantly shorter than the lifetime of the solar nebula at the optimal temperatures around 550 K. These authors thus concluded that $CH_4$-rich satellitesimals could have formed in the catalytically active region of the subnebula and thus may have played an important role in the origin of Titan's atmosphere. Despite the fact that these experiments suggest that Titan's atmospheric methane could have been produced in the gas phase of Saturn's subnebula, they are not found to be consistent with the high D/H ratio measured of this molecule, which is found to be ~6 times the protosolar value (Bézard *et al.*, 2007). On the contrary, the D/H in $CH_4$ produced in Saturn's subnebula should be very close to the protosolar value (Mousis *et al.* 2002), thus invalidating the idea that this molecule was formed from CO and protosolar $H_2$ in Saturn's subnebula. The only scenario that remains consistent with this constraint is that $CH_4$ was originally formed in the Interstellar Medium and subsquently accreted in Titan's building blocks in the protosolar nebula (Mousis *et al.*, 2009).

Interestingly, Enceladus is a good candidate for the existence of serpentinization and FTT reactions in its interior. The plumes of vapor and water ice particles rich in sodium salts erupting from the south polar region of Enceladus suggest the presence of a liquid water reservoir below the crust (Potsberg *et al.*, 2009, 2011; Waite *et al.*, 2009). Moreover, the recent discovery of silicate nanoparticles derived from the plumes by the Cassini spacecraft indicates the presence of rocks in contact with Enceladus' ocean (Hsu *et al.*, 2014). These observations are complimented by the observations of $CH_4$ in Enceladus' plumes (Waite *et al.*, 2009). According to laboratory experiments (Sloan and Koh, 2008; Vu and Choukroun, 2015), methane clathration should be very efficient in the internal ocean and suggests that methane should be at very low concentrations in the plume. A simple way around this would be to have active production of methane via water-rock interactions (Bouquet et al., 2015). This has motivated a new interest in conducting serpentinization experiments in the appropriate temperature and pressure ranges (Sekine *et al.*,

2014).

## 5. Conclusion and future prospects

Methane is the simplest hydrocarbon and can be observed in a wide variety of extraterrestrial environments, including interstellar space and atmospheres of gas giants and Titan. Both methane formation processes and further reactions from methane have been studied by way of theoretical calculations and simulation experiments.

To simulate interstellar gaseous reactions, experiments are performed under low pressure, including those done with CRESU. In addition to the gas phase, reactions on the surfaces of ice mantles of interstellar dust particles (ISDs) should not be ignored. A possible formation pathway of methane is the addition of H atoms to C atoms on dust surfaces. This pathway has not been unambiguously established experimentally, but the formation of $CH_3OH$ from CO via HCHO has been experimentally confirmed.

Another series of experiments involve the synthesis of complex organic compounds from possible molecules in the ice mantles of ISDs. Ultraviolet light and cosmic rays are two possible energy sources, and complex molecules including amino acid precursors were formed from such icy mixtures. In these experiments, CO and/or $CH_3OH$ were mainly used as carbon sources, and amino acid precursors were formed. When methane was used as a sole carbon source, however, amino acid precursors were not formed by UV irradiation. The role of methane in the ice mantles of ISDs in the formation of complex molecules should be examined in future experiments.

In the mid-20[th] century, a strongly reducing primitive Earth atmosphere containing a high concentration of methane was postulated, and a great number of experiments were conducted. The presence of such an atmosphere was disproved later, but it was suggested that methane presented as a minor constituent. Even now, a supply of a small amount of methane is observed in submarine hydrothermal systems. Opportunities for those studying chemical evolution in submarine hydrothermal systems include demonstrating the importance of quenching key chemicals processed in hydrothermal systems are expected. In addition, it will be important to show the roles of methane in reactions in flow reactors.

Titan, which has a dense atmosphere that contains methane, is considered as a natural laboratory of chemical evolution.   A wide variety of laboratory experiments have been performed to simulate the higher or lower atmosphere of Titan with various energies such as solar UV, electrons in a saturnian magnetosphere, and cosmic rays.   It has been shown that complex solid organics, sometimes referred to as *tholins*, were formed in these simulation experiments, which corresponded to haze in Titan's atmosphere as observed by the Cassini-Huygens mission. Another important topic concerning methane with regard to Titan is the discovery of liquid ethane-methane lakes on Titan.   These tholins, when discussed in the context of a possible generation of life in Titan's liquidosphere, suggest future simulation experiments that include possible chemical evolution studies, which would be important to design future Titan missions.

Not only Titan, but other icy bodies as well such as Enceladus and Pluto, would be good targets with which to further our understanding of the diversified methane chemistry in space, and experiments simulating environments of these bodies are expected.

There have been a great number of experiments implemented to investigate the chemistry of methane and related molecules in various environments.   However, there has been little collaboration among research teams involved in such experiments.   For example, experiments simulating surface reactions on ISD ice mantles have been conducted mostly independently from those modeling photochemical / radiochemical reactions in those mantles. Further interdisciplinary collaborative work among different research groups could have a significant impact on this aspect of astrobiology. An additional key goal is the formulation of an overall synthesis scenario from interstellar molecules to prebiotic molecules, which might occur in terrestrial submarine hydrothermal systems or Titan's hydrocarbon lakes.

**Acknowledgements**

funded by the International Space Science Institute (ISSI). Therefore, a certain focus will be on the chemistry of methane and hydrocarbons.

The authors gratefully acknowledge support from the International Space Science Institute (ISSI) for our team "The Methane Balance - Formation and Destruction Processes on Planets, their Satellites and in the Interstellar Medium", Team ID 193.   O.M. acknowledges support from CNES. This work has been partly carried out thanks to the support of the A*MIDEX project (n°ANR-11-IDEX-0001-02) funded by the "Investissements d'Avenir" French Government program, managed by the French National Research Agency (ANR). M.E.P. acknowledges the support by the Italian Ministero dell'Istruzione, dell' Università e della Ricerca (MIUR) through the grant Progetti Premiali 2012 - iALMA (CUP C52I13000140001). NC acknowledges the support of the European Research Council (ERC Starting Grant PRIMCHEM, grant agreement n°636829). This work was also supported by the NASA NExSS program via grants NNX13ZDA017C and NNX15AD94G.

No competing financial interests exist.

FIGURE CAPTIONS

FIG. 1. Basic scheme of a selected ion flow tube apparatus, which consists of an ion source, a quadrupole mass filter to isolate a given ion reactant, the reaction zone where neutral species are introduced, and a quadrupole filter/channeltron product analyzer (from Smith and Adams 1988).

FIG. 2. Schematics of a Flowing afterglow apparatus. Ions (and electrons) are produced in the microwave discharges. By means of introduction of a reactant secondary, ions can be produced. With the movable Langmuir probe, the decay of electrons through recombination reaction can be followed. Taken from Larsson et al. 2012.

FIG. 3. The CRYRING magnetic storage ring. Ions are produced in the MINIS ion source and injected into the ring *via* a radiofrequency quadrupole. After acceleration in the ring by a radiofrequency field the ion beam is merged with an electron beam in the electron cooler. Neutral products of ion-electron reactions are unaffected by the magnetic field and leave the ring tangentially and can be detected by the surface barrier detector. Taken from the work of Larsson et al. 2012.

FIG. 4. Schematic of the CRESU apparatus. Reactants are produced from the precursor in the moveable    reservoir and enter the vacuum chamber in a supersonic flow of the carrier gas through the Laval nozzle. The decay of the reactant is followed by means of Laser induced fluorescence by the photomultiplier (PMT). The variable distance between the reservoir and the detector allows for observation of the intensity of the reactant in the supersonic flow at different flow times. Taken from the work of Smith (2006).

FIG. 5.   Normalized column density of $CH_4$ as a function of irradiation dose (eV/16u) after ion bombardment of pure $CH_4$ and a $H_2O:CH_4$=4:1 mixture at 12 K with 30 keV $He^+$.

FIG. 6. Top panel: Normalized column density of $CH_3OH$ as a function of irradiation dose (eV/16u) after ion bombardment of pure $CH_3OH$ and two mixtures, $CO:CH_3OH$ and $N_2:CH_3OH$, at 16 K with 200 keV $H^+$.

Bottom panel: Column density of $CH_4$ formed at 16 K after ion bombardment (200 keV $H^+$) of the same $CH_3OH$-rich ice mixtures divided by the column density of $CH_3OH$ at the same dose.

FIG. 7. Comparison between observed and laboratory spectra.   Data points are ISO observations.   In each panel the thick red line is the best-fit to the observed data points obtained by a linear combination of two laboratory spectra. The thin lines are the laboratory spectra scaled by the coefficient given by the fit procedure.

FIG. 8. Schematic introduction of the CAPS and INMS data sets that were used to study the ion-neutral chemistry producing complex organic compounds in Titan's upper atmosphere. Panel A shows the neutral (upper) and ion (lower) mass spectra from 1 to 100 u in Ttian's upper atmosphere measured by INMS. Panel B shows the mass spectra of Crary et al. (2009) taken by CAPS IBS and indicating positive ions that contain PAHs and nitrogen substituted aromatics that extend to >250 u. Panel C is a negative ion spectrum from the work of Coates et al. (2009) indicating that the ion neutral chemistry also involves negative ions with masses of over 5000 u.

FIG. 9.   The pressure regime of tholin laboratory experiments versus the in situ pressure conditions measured by Cassini-Huygens. Ion neutral organic fomation processes dominate at low pressures, whereas radical three body stabilized reactions predominate at higher pressures.